\documentclass[
    ,final            % use final for the camera ready runs
  ]
  {aipproc}

\layoutstyle{8x11double}

\begin{document}

\title{An Update on the 0Z Project}


\author{J.~G.~Cohen}{
  address={California Institute of Technology}
}

\author{N.~Christlieb}{
  address={Uppsala University, Sweden}
}

\author{A.~McWilliam}{
  address={Carnegie Observatories of Washington}
}

\author{S.~Shectman}{
  address={Carnegie Observatories of Washington}
}

\author{I.~Thompson}{
  address={Carnegie Observatories of Washington}
}

\begin{abstract}

We give an update on our 0Z Survey  to find more extremely metal poor 
(EMP) stars with [Fe/H] $< -3$~dex through mining the database of
the Hamburg/ESO Survey.  We present the most extreme such stars
we have found from $\sim 1550$ moderate resolution follow up spectra.
One of these, HE1424$-$0241, has highly anomalous abundance ratios
not seen in any previously known halo giant, 
with very deficient Si, moderately deficient Ca and Ti, 
highly enhanced Mn and Co,
and low C, all with respect to Fe.  We suggest a SNII
where the nucleosynthetic yield for explosive $\alpha-$burning nuclei
was very low compared to that for the hydrostatic $\alpha-$burning element Mg,
which is normal in this star relative to Fe.  A second, less
extreme, outlier star with high [Sc/Fe] has also been found.

We examine the extremely metal-poor tail of the HES metallicity
distribution function (MDF).  We suggest on the basis of
comparison of our high resolution detailed abundance analyses
with [Fe/H](HES) for stars in our sample that the MDF
inferred from follow up spectra of the HES sample
of candidate EMP stars is heavily contaminated
for [Fe/H](HES) $< -3$~dex; many of the supposed EMP
stars below that metallicity are of substantially higher
Fe-metallicity, including most of the very C-rich stars, 
or are spurious objects.

\end{abstract}

\maketitle

%%%%%%%%%%%%%%%%%%%%%%%%%%%%%%%%%%%%%%%%%%%%
%% MAINMATTER
%%%%%%%%%%%%%%%%%%%%%%%%%%%%%%%%%%%%%%%%%%%%

\section{Introduction}

Extremely metal poor (EMP) stars provide important clues to the chemical
history of our Galaxy, the role and type of early SN, the
mode of star formation in the proto-Milky Way, and the formation
of the Galactic halo.  The sample of known EMP
stars is summarized by \cite{beers05}.  They
compiled a list of the key properties of
the 12 stars identified up to that time with [Fe/H] 
$\leq -3.5$~dex, 7 of which are EMP giants and subgiants within
the range of $T_{eff}$ between 4900 and 5650~K. 

Our 0Z project has the goal of increasing the sample of such stars
through data mining of
the  Hamburg/ESO Survey \citep{wis00}.  This is an 
objective prism survey from which it is
possible to efficiently select QSOs \citep{wis00} as well
as a variety of interesting
stellar objects, among them EMP stars 
\citep{christlieb03}.

Our 0Z project has been systematically searching the 
database of the HES for this purpose over the past five years.
We have just published \citep{cohen08}, and we
present at this conference, a sample of new EMP giants 
with $T_{eff} < 6000$~K
and [Fe/H] $\leq -3.5$~dex which substantially increases the
number of such stars known.

\section{Outliers}

Our 0Z project has obtained moderate resolution 
follow up spectra
of $\sim$1550 candidate EMP stars selected from the HES.
These follow up spectra were processed through the
code described in \cite{beers99}, which
is essentially identical to that
used by the HK Survey\citep{beers92} until recently; 
the latest updates to the algorithm
as used by the HK Survey are described in \cite{rossi05}.  
The algorithm uses
the strength of H$\delta$ and of the 3933~\AA\ Ca~II line
to assign a metallicity to each star, which we denote
as [Fe/H](HES). 
(The SEGUE survey uses an even more sophisticated algorithm
taking advantage of the uniform wide wavelength coverage
of the SDSSII spectra and multi-color photometry;
a preliminary description is given in \cite{segue}.)
We attempt to observe all stars with [Fe/H](HES) $< -2.9$~dex 
as inferred from
these moderate resolution follow up spectra at high resolution.

We present the results of detailed abundance analyses based on high resolution
and high signal-to-noise spectra of eight extremely metal poor (EMP)
stars with [Fe/H] $\leq -3.5$~dex, four of which are new.
Only stars with $4900 < T_{eff} < 5650$~K are included.

Past work on EMP stars, e.g. \cite{cohen04}, \cite{cayrel04}
and \cite{arnone05},
has emphasized the small scatter at a fixed
[Fe/H] for trends of element abundance ratios [X/Fe].  The
$\sigma$ about the mean at each [Fe/H] appears to be  consistent
with very little intrinsic scatter, just the observational errors
contributing.  This applies only to C-normal metal-poor stars,
and to elements between Na and the end of the Fe-peak.  The light
elements Li, Be, C, N, and O 
in EMP first ascent RGB stars may be affected by mixing of internally processed
material,
while EMP AGB stars may have Na and Al affected as well \citep{spite05}.
We established in \citep{cohen03} that the neutron capture
elements are decoupled from the Fe-peak elements;
the site or conditions of their production, presumably via the 
$r$ or $s-$process, must be different from 
that of the Fe-peak elements.

The most interesting thing we have discovered recently
is the existence of outliers in these relations among
EMP stars.  The most sensational
case we have found, for which a brief description was
given in \cite{cohen07} with a full study in \cite{cohen08},
is HE1424$-$0241.  This star is the most metal poor star 
in our sample, with [Fe/H] $\sim -4$~dex.
It has highly anomalous abundance ratios unlike those of any other
known EMP giant, with very low Si, Ca and Ti relative to Fe,
and enhanced Mn and Co, again relative to Fe. 
Figs.~1 and 2 illustrate the uniqueness of this star
among other halo giants
with regard to its Ca and Si abundance ratios with
respect to Fe.
Only (low) upper limits
for C and N can be derived from the non-detection of the CH
and NH molecular bands.

Si, Ca and Ti are formed via explosive $\alpha-$burning
while Mg, which is normal with respect to Fe in HE1424$-$0241, is formed
via hydrostatic $\alpha-$burning.  A much smaller
separation of abundance ratios ($\sim 0.2$~dex) for elements between
these two types of $\alpha-$burning was seen
by \cite{fulbright07} in the Galactic bulge, 
but that of HE1424$-$0241 is much more extreme,
and cannot be explained with recent SN nucleosynthesis yields. 
We suggest a peculiar SNII produced the chemical inventory
of this star.
HE0132$-$2429, another sample star, has 
excesses of N and Sc  with respect to Fe.

%%%%%%%%%%%%%%%%%%%%%%%%%%%%%%%%%%%%%%%%%%%%
%% Sample figure:
%%
%% The option [height=...] scales the picture to the given height,
%% without it it would be printed at its nominal size
%%%%%%%%%%%%%%%%%%%%%%%%%%%%%%%%%%%%%%%%%%%%

\begin{figure}
  \includegraphics[height=.4\textheight]{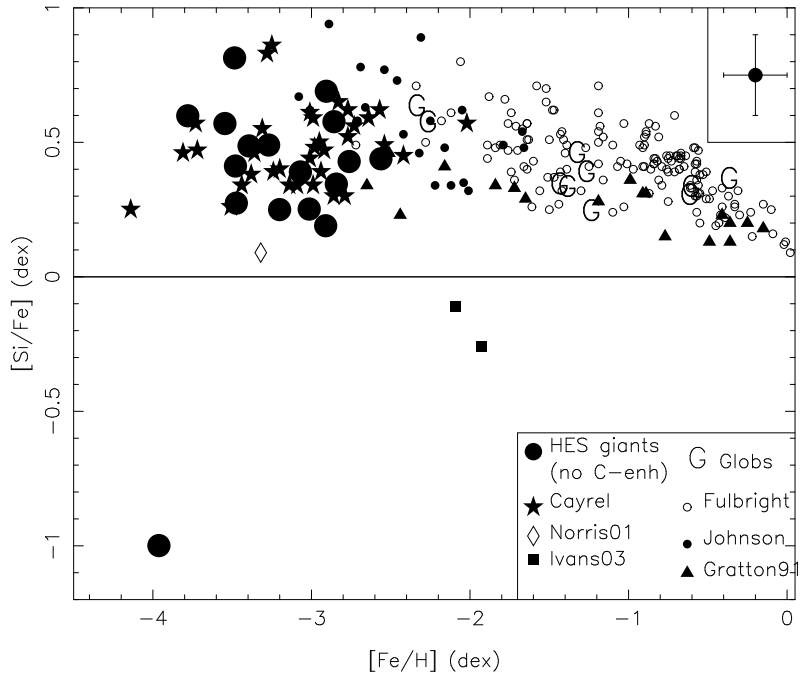}
  \caption{[Si/Fe] is shown for all of the candidate EMP stars
with HIRES spectra analyzed by the 0Z project to date, including
the present sample.  C-rich stars are not shown. The solid horizontal line denotes
the Solar ratio.  The plot includes well studied Galactic globular clusters,
mostly from analyses by J.~Cohen and her collaborators, as well as
samples of halo field stars from the sources indicated on the symbol key
in the lower right of the figure.
Note the highly anomalous position of HE1424$-$0241, the only star
with [Si/Fe] $<< 0$~dex.}
\end{figure}
% Figure file copied: 
% /scr2/jlc/hamburg_survey/paper_lowest_2006/plots/si_feh_all_27aug2007.ps

\begin{figure}
  \includegraphics[height=.4\textheight]{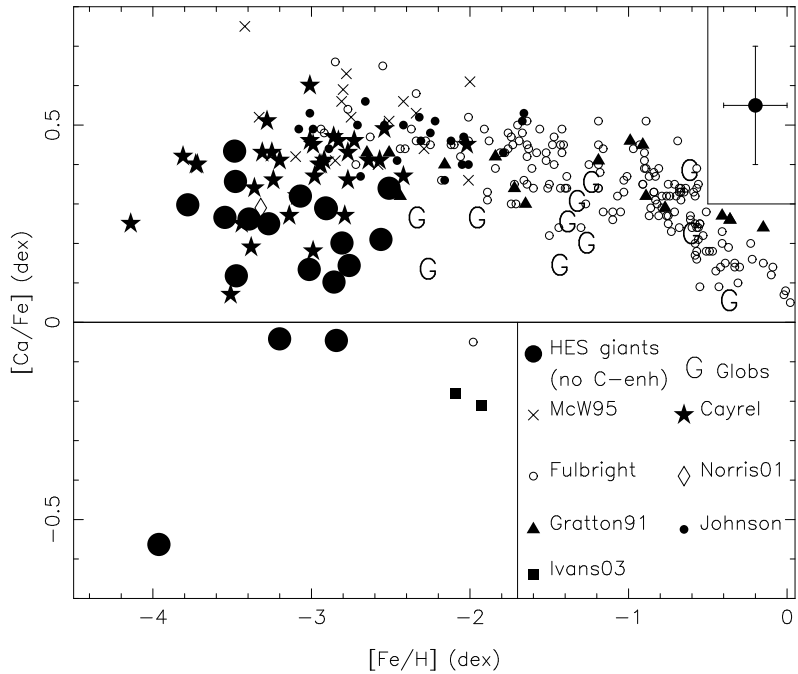}
  \caption{The same as Fig.~1 for [Ca/Fe].}
\end{figure}
% Figure file copied:
% /scr2/jlc/hamburg_survey/paper_lowest_2006/plots/ca_feh_all_27aug2007.ps

The strong outliers in abundance ratios among
the Fe-peak elements in these C-normal stars,
not found at somewhat higher metallicities
([Fe/H] $\sim -3$ dex),  
are definitely real.
The anomalies discussed
here cannot be due to observational or analysis errors.
There are two HIRES spectra for each of these two stars.
The effects are very large, and are measured relative
to the many other EMP stars we have analyzed in a homogeneous
manner with identical codes and procedures.

They suggest that at such low metallicities
we are beginning to see the  anticipated and
long sought stochastic effects of individual supernova events
contributing to the Fe-peak material within a single star.
   
Another of our recent results 
\citep[see][]{cohen08} comes from a
detailed comparison of the analysis procedures
adopted by our 0Z project compared to those of the First Stars VLT
Large Project.  We find a systematic difference for [Fe/H] of
$\sim$0.3 dex, our values always being higher, largely due
to differences in adopted reddenings, temperature scales,
and Fe solar abundance between the two projects.

\section{The Metallicity Distribution Function of the HES}

A preliminary metallicity distribution function for the HES
was published in  \cite{beers05}.  We examine here the likely
validity of this result, focusing on the most extreme
of the EMP stars.  We ignore issues
of incompleteness, which are currently being studied by
T.~Sch\"orck and N.~Christlieb, and which while small at [Fe/H] $< 3$~dex, 
become quite large by $-2$~dex.

Our 0Z project has obtained moderate resolution 
follow up spectra
of $\sim$1550 candidate EMP stars selected from the HES.
We have attempted to observe all stars with 
[Fe/H](HES) $< -2.9$~dex  as inferred from these
follow up spectra at high resolution.
We list below the 12 objects from the 0Z project for
which the resulting [Fe/H](HES) was less than
$-3.5$~dex. (The ID is truncated for stars whose analyses 
we have not yet published.)

\begin{table}
\begin{tabular}{lrrl}
\hline
\tablehead{1}{r}{b}{Star Name}
  & \tablehead{1}{r}{b}{[Fe/H](HES)}
  & \tablehead{1}{r}{b}{[Fe/H](HIRES)}
  & \tablehead{1}{r}{b}{Comments}   \\
\hline
HE1159$-$0525 &  $-$3.52   &     ?? &     Extreme C$-$star, [Fe/H](HES) prob too low \\
%  &    &    &   [Fe/H](HES) prob too low \\
HE1150$-$0428 &  $-$3.57    &    $-$3.27 &  C$-$star \\
HE2209     &   $-$3.57   &  $-$3.47  & MIKE spectrum \\
HE1031$-$0020 &  $-$3.61 &    $-$2.88 & C$-$star \\
HE0122     &   $-$3.63   &     $-$2.80 \\
HE0305$-$5442 &  $-$3.68 &       $-$3.56 & CS22968$-$014 rediscovery, [Fe/H] from 
   Cayrel et al (2004) \\
%    &    &    &  Cayrel et al (2004) \\
HE0313    &    $-$3.72   &     $-$3.63 &  MIKE spectrum \\
HE0911$-$0512  & $-$3.84  &  ...  &    M dwarf \\
HE2323$-$0256  & $-$3.96 &      $-$3.78 &  CS22949$-$037 rediscovery \\
HE1432   &   $-$4.43  &      $-$2.51 \\
HE0208$-$5335 &  $-$4.14     &  ... &    QSO (Ian Thompson) \\
HE1030+0137 &  $-$4.37  & ... &    M dwarf \\
\hline
\end{tabular}
\caption{Status of Stars From Raw 0Z Survey With [Fe/H](HES) $< -3.5$~dex}
\label{table_list}
\end{table}

Since these are in principle stars of great interest, the
follow up spectra were scrutinized with care.
Three of the 12 were eliminated in that way; two are M dwarfs and one
is a  QSO.  Each
object that survived (with one exception) has now
been observed at high resolution with either
HIRES at Keck or MIKE at Magellan. 

Of  the three objects predicted to have [Fe/H] $< -4$~dex,
none have [Fe/H](HIRES) $< -2.8$~dex.  This leaves
8 of the 12 objects.
Three of these are C-stars, two of them have
[Fe/H](HIRES) $> -3.3$~dex, the third is the only
one that does not yet have a high resolution spectrum.
It is, however, an C-star with extremely strong CH, and
therefore is very likely to have [Fe/H] $> -3.5$~dex
based on our earlier work described in \cite{cohen05}.
We assume for present purposes that this is the case;
a HIRES spectrum to be obtained shortly should confirm this.

Two of the stars listed in the table
are rediscoveries of previously known EMP stars found by the
HK Survey\citep{beers92}.  In the end, after all this
effort, we have found only two new stars with
[Fe/H](HIRES) $< -3.5$~dex, out of $\sim$1550
follow up spectra.

Genuine EMP stars are very rare, and the EMP
tail of the HES metallicity distribution function is heavily
contaminated with
stars which because of observational or analysis errors have landed
up there, as well as with C-stars whose metallicity has been
underestimated \citep[see][]{cohen05} and with some spurious
objects.

Algorithms currently in use with SEGUE do not attempt to
assign a metallicity  based on H$\delta$ and the Ca~II K line
to very C-rich stars. 
This is a big step forward.  However,
the extreme low metallicity end of any MDF generated from HES follow up
spectra is likely to be heavily contaminated with
additional higher metallicity objects which through various issues,
including observational errors,
end up getting assigned a spurious very low [Fe/H](HES).

We now have HIRES observations in hand for essentially
all of the 16 stars from our 0Z project with $-3.5 <$ [Fe/H](HES) $< -3.3$~dex.
Of the six analyzed thus far, only three have
[Fe/H](HIRES) $< -2.8$~dex. 
We are in the process of analyzing the rest of these to see what
fraction of those in the regime 
of [Fe/H](HES) between $-3.5$ and $-3.3$~dex
are genuine.   In the meantime, we urge caution in looking
at the low-metallicity tail of any preliminary MDF from the HES
in which the metallicities of
   stars at [Fe/H] $< -3.0$~dex are based on moderate-resolution
   spectra rather than high-resolution spectroscopy.

%%%%%%%%%%%%%%%%%%%%%%%%%%%%%%%%%%%%%%%%%%%%%%%%
%% BACKMATTER
%%%%%%%%%%%%%%%%%%%%%%%%%%%%%%%%%%%%%%%%%%%%%%%%

\begin{theacknowledgments}

Based in part on observations obtained at the
W.M. Keck Observatory, which is operated jointly by the California 
Institute of Technology, the University of California, and the
National Aeronautics and Space Administration.
Some of the data presented herein were obtained at the Palomar
Observatory.
We are grateful to the many people  
who have worked to make the Keck Telescope and HIRES  
a reality and to operate and maintain the Keck Observatory. 
J.G.C. is grateful to NSF grant AST-0507219  for partial support.
      N.C. is a Research Fellow of the Royal Swedish Academy of
      Sciences supported by a grant from the Knut and Alice
      Wallenberg Foundation. He also acknowledges financial
      support from Deutsche Forschungsgemeinschaft through grants
      Ch~214/3 and Re~353/44.  We thank Tim Beers for providing
      us with [Fe/H](HES) values using the most recent version of his codes.
\end{theacknowledgments}

\end{document}